\documentclass[12pt]{article}
\usepackage{graphicx}
\usepackage{amssymb,amsmath,amsfonts,palatino,amsthm}
\usepackage{amssymb}
\usepackage{epstopdf}
\newcommand{\beq}{\begin{equation}}

\def\Z#1{_{\lower2pt\hbox{$\scriptstyle#1$}}}

\begin{document}

\title{The warping of extra spaces accelerates the expansion of
the universe\footnote{Essay written for the Gravity Research
Foundation 2010 Awards for Essays on Gravitation; Selected for an
``Honorable mention" 15 May 2010.}}
\author{Ishwaree P. Neupane\footnote{E-mail: ishwaree.neupane@canterbury.ac.nz}
\\
Department of Physics and Astronomy, University of Canterbury\\
Private Bag 4800, 8041 Christchurch, New Zealand}
\date{March 31, 2010}
\maketitle

\begin{abstract}

\noindent Generic cosmological models derived from higher
dimensional theories with warped extra dimensions have a nonzero
cosmological constant-like term induced on the 3+1 space-time, or
a physical 3-brane. In the scenario where this 3+1 space-time is
an inflating de Sitter ``brane" embedded in a higher-dimensional
space-time, described by warped geometry, the 4D cosmological term
is determined in terms of two length scales: one is a scale
associated with the size of extra dimension(s) and the other is a
scale associated with the warping of extra space(s). The existence
of this term in four dimensions provides a tantalizing possibility
of explaining the observed accelerating expansion of the universe
from fundamental theories of gravity, e.g. string theory.

\end{abstract}
\newpage


Physicists today are faced with a number of mysteries about the
universe which have called for elegant and straightforward
explanations. One major mystery uncovered nearly 12 years ago from
observations of distant supernovae~\cite{supernovae} is that the
expansion of the present-day universe is accelerating rather than
slowing down.

In the standard model of cosmology, first, we introduce a
cosmological constant $\Lambda$, which has the same effect as an
intrinsic energy density of the vacuum (and an associated negative
pressure). It is used to balance the total energy budget and
explain the accelerating expansion observed in the more recent
history of our universe. Second, we put in dark matter, which
plays a key role in the formation of cosmic structure. Third, we
demand the occurrence of cosmic inflation at any early epoch -- a
period of exponential expansion that explains why the universe is
smooth on the largest scales.

The Einstein's cosmological constant appears to be the most
parsimonious solution to a late time cosmic
acceleration~\cite{Online}. Yet, a canonical picture of the
standard model of cosmology as above raises some conceptual
issues, such as, what is the origin of a cosmological term in
Einstein's theory of general relativity (GR)? Is gravity purely
geometrical as Einstein thought, or is there more to it (such as,
extra dimensions, scalar fields)? What made our universe so big?
This last question can be addressed by invoking a period of cosmic
inflation, but again we do not know the physical origin of
gravitationally repulsive energy driving inflation. Also, there is
no known natural way to derive a tiny cosmological constant - that
the observations favour - from fundamental theories of gravity.

Given that the physics in higher spacetime dimensions defines a
framework of low-energy four-dimensional field theory and a unique
answer to the gravitational vacuum energy density in today's
universe, there can be a unique prediction for the cosmological
constant. I want to argue, in this essay, that the cosmological
constant of our spacetime may well be determined by two
parameters: one is a scale related to the size of extra dimensions
and the other is a scale related to the warping of extra
dimension(s). This essay, building on some new ideas, provides a
key conceptual consolidation.

Physicists have long sought a theory of quantum gravity that would
unify all the principles of physics in a pleasing mathematical
formalism. The history of 20th century physics has itself been a
struggle to find a way to unite general relativity and quantum
mechanics. One of the ways in which physicists have tried to unify
gravity with the other forces is to create theories that allow for
many more spatial dimensions. A serious motivation for considering
additional dimensions came from string theory, which in turn is
inspired by the failure of classical (Einstein) gravity to work at
very short distance scales.

In the early 1980s, certain ten- and eleven-dimensional
supergraviy models - in the low energy limits of 10D string theory
and 11D M-theory - were realized as profound theoretical tools in
connecting quantum mechanics with general relativity. The major
achievement in the mid 1990s was a series of new understandings
about the role of localized sources such as `branes' and
fluxes~\cite{Polchinski}. These ingredients in string/M-theory
have helped us to learn several new ideas in physics, including
the localization of gravity on a brane embedded in a warped
five-dimensional anti-de Sitter space~\cite{RS}, gravity - gauge
theory dualities~\cite{Malda} and methods of constructing de
Sitter vacua in string theory~\cite{KKLT}.

Despite some novelties, it is not straightforward to explain
cosmic acceleration directly through a standard compactification
of ten- or eleven-dimensional supergravity models. There is a
`no-go' theorem, due to Gibbons~\cite{Gibbons-84}, Maldacena and
Nunez~\cite{Malda-Nunez} and many others~\cite{GKP}, which
basically asserts that if we compactify any string-derived
supergravity model on a smooth compact manifold, then it would be
difficult to obtain an inflationary cosmology as a background
solution of 10D or 11D classical supergravities.

Since the universe appeared to be both past and future de Sitter
(albeit with vastly differing vacuum energies) this would seem to
be a problem. In turn, in recent years attempts have been made
around this particular no-go result. The original no-go theorem
assumes time independence of the internal space, and so one could
look for time-dependent solutions. Following this intuition, the
author and many others~\cite{TW,Ish03c} have constructed varieties
of time-dependent compactifications which describe an accelerating
universe. In several examples studied in the literature, with
maximally symmetric extra dimensions, solutions to D-dimensional
Einstein equations (with $D\ge 5$) only support a transiently
accelerating universe with time-dependent metric moduli.

Indeed, almost all solutions found using time-dependent metric
moduli have the property that as one moves to the minimum of the
potential the extra dimensions grow or shrink slowly or even
stabilise in a few specific cases~\cite{Ish07a}, resulting into an
exponential type potential. One can now imagine "bouncing" the
universe off this potential. Albeit for a brief interval, the
energy is dominated by the potential term of metric moduli, or
scalar fields, and the universe undergoes a transient period of
cosmic acceleration~\cite{Ish03c}.

To that end, we find interest in an alternative scenario where the
extra dimensions are warped but time-independent. One of such
proposals is the five-dimensional `warped' braneworld model
proposed by Randall and Sundrum~\cite{RS}. Their proposal raises
the possibility that at least one of the extra dimensions
postulated by string theory could be large enough to have
experimental implications, including that in particle physics.

If we learned one thing from higher dimensional theories of
gravity, including string theory, it is that extra dimensions play
major tricks in a spacetime that is dynamical. In what follows, I
use very little of the full formalism that has been developed to
describe string theory, rather motivate the readers to one basic
fact. Gravity, Einstein asserted, is caused by a warping of space
and time - or, in a language many physicists prefer, by a warping
of spacetime. Although Einstein's original assertion was based on
dynamics in a stationary and in a four-dimensional spacetime, the
same is true in a dynamical background and with an arbitrary
number of spacetime dimensions.

One quite simple, but equally robust, observation is that, in
spacetime dimensions $D\ge 5$, with $(D-4)$ warped extra
dimensions, a four-dimensional observer would measure a nonzero
positive cosmological constant, which is induced not by a
D-dimensional cosmological term but by the curvature related to
the expansion of a physical 3+1 spacetime. This basic idea can be
illustrated with a five-dimensional `warped metric', maintaining
the usual four-dimensional Poincar\'e symmetry,
\begin{equation}\label{sol-5Da}
ds_5^2 = e^{2A(y)} \left(\hat{g}_{\mu\nu} dx^\mu dx^\nu + \rho^2\,
{dz}^2\right), \quad A(z)=-\mu |z|,
\end{equation}
where $x^\mu$ are the usual coordinates ($\mu, \nu=0,1, 2, 3$),
$\rho$ is a length scale associated with the size of extra space,
$e^{A(y)}\equiv e^{-\mu|z|}$ is the warp factor as a function of
the fifth dimension $z$. Indeed, the problem of the vacuum energy
density or cosmological constant - why it is extremely small by
particle physics standard - arises only in a dynamical spacetime.
What it means is that the usual 4D line element must be
time-dependent, such as
\begin{equation}\label{FRW}
\hat{g}_{\mu\nu} dx^\mu dx^\nu =   -dt^2+ a^2(t)\left[\frac{
dr^2}{1-k r^2}+ r^2 (d\theta^2+\sin^2\theta d\phi^2)\right],
\end{equation}
where the curvature constant $k=0, + 1$ and $-1$, respectively,
for flat, closed and open universes. With (\ref{FRW}), the metric
(\ref{sol-5Da}) becomes an exact solution of 5D Einstein
equations, following from
\begin{equation}
S \propto  \int d^5{x} \sqrt{-g_5} R_5,
\end{equation}
when the scale factor of the universe $a(t)$ takes the form
\begin{equation}\label{main-sol-scale1}
a(t)= \frac{1}{2} \exp\left(\frac{\mu t}{\rho} \right) +\frac{{k}
\rho^2}{2 \mu^2}\, \exp\left( - \frac{\mu t}{\rho}\right),
\end{equation}
where $\mu$ is some constant (with inverse length dimension). When
$\mu=0$, it is mandatory to take $k=0$ and also introduce a 5D
cosmological term, or a source in the 5D bulk spacetime, which
corresponds to a static configuration studied in Ref.~\cite{RS}.

Especially, in the $k=-1$ case, there is a big-bang type
singularity at
\begin{equation}
t=\frac{\rho}{2\mu} \ln \frac{\mu^2}{\rho^2},
\end{equation}
whereas in closed and spatially flat universes, the scale factor
is nonzero. This is desirable because the generic singularity of
time-dependent solutions of general relativity is generally
unacceptable - as it cannot have any quantum interpretation. More
remarkable is the fact that the universe naturally inflates when
$\mu>0$ or the warp factor is not constant.

In the discussion above, I assumed for simplicity that the 5D bulk
cosmological term is zero, but a similar solution exists when one
modifies the 5D classical action as
\begin{equation}
S \propto  \int d^5{x} \sqrt{-g_5} \left( R_5- 2 \Lambda_5\right).
\end{equation}
The only difference now is that the warp factor is given by
\begin{equation}
A(z)= \ln (24\mu^2)-\ln B(z), \quad {\rm where} \quad B(z) =
24\mu^2 e^{\mu |z|} + \Lambda_5 \rho^2 e^{-\mu|z|},
\end{equation}
which reduces to the earlier result when $\Lambda_5$ is set to
zero. In the above I replaced $z$ by mod $z$, or imposed a ${\cal
Z}_2$ symmetry at $z=0$; the reason was that in a physical
scenario, one has to solve the D-dimensional Einstein equations
not just for $|z|>0$ but also at the position of a physical
3-brane or the 4D hypersurface at $|z|=0$, which can in principle
have a nonzero 3-brane tension $T_3$, described by the action
\begin{equation}\label{5d-main}
S = \frac{M_5^3}{2} \int d^5{x} \sqrt{-g_5} \left( R_5- 2
\Lambda_5\right)+ \int d^4{x} \sqrt{-g_b} (-T_3),
\end{equation}
where $M_5$ is the 5D Planck mass. The 5D Einstein equations
\begin{equation}
G_{AB} = -\Lambda_5 g_{AB} - \frac{T_3}{M_5^3}
\frac{\sqrt{-g_b}}{-g_5} g_{\mu\nu}^b \delta_A^\mu \delta_B^\nu
\delta(z),
\end{equation}
following from the action (\ref{5d-main}), are explicitly solved
when
\begin{equation}
T_3 = \frac{6\mu M_5^3}{\rho^2}\left(1-\frac{\Lambda_5
\rho^2}{24}\right).
\end{equation}
It is assumed that $\Lambda_5<0$ or $\Lambda_5=0$, which both lead
to a positive brane tension. In the $\Lambda_5>0$ case we can
actually relax the $Z_2$ symmetry, which still leads to a finite
4D Newton constant, but we then need to satisfy $\Lambda_5<
24/\rho^2$ so as to get a positive brane tension.

An important observation here is that the brane tension is induced
by the curvature related to the expansion of $3+1$ spacetime
alone, which is determined by the slope of the warp factor. In a
wide class of warped compactifications, which include string
theory, this suggestion has startling implications.

First, the physical universe inflates naturally when the warping
of extra space is nonzero. This happens because the warping of
extra space(s) generates in the four-dimensional effective theory
a cosmological constant-like term $\Lambda_4$. To show this, we
focus here on the $R_5-2\Lambda_5$ term from which we derive the
scale of gravitational interactions:
\begin{eqnarray}\label{effective-S}
S_{\rm eff} & \supset &  \frac{ M_5^3\rho }{2}  \int e^{3A(z)} dz
\int d^4{x} \sqrt{-\hat{g}\Z{4}} \left(\hat{R}_4 -
\frac{12}{\rho^2} (\partial_z{A})^2- \frac{8}{\rho^2}
\partial_z^2 A -2\Lambda_5 e^{2A(z)} \right)
\nonumber \\
 & \supset & \frac{ M_5^3\rho }{2}  \int
\left(\frac{24\mu^2}{B(z)}\right)^3 dz \int d^4{x}
\sqrt{-\hat{g}\Z{4}} \left(\hat{R}_4 - \frac{12 \mu^2}{\rho^2} +
\frac{768\Lambda_5 \mu^4}{B^2(z)} \right).
\end{eqnarray}
When $\Lambda_5= 0$, $\Lambda_4 (\equiv 6 \mu^2/\rho^2)$ is
constant, whereas, with $|\Lambda_5|> 0$, it is position
dependent.

Second, as advertised above, the vacuum energy density is uniquely
determined in terms of two length scales: in the $\Lambda_5=0$
case, the two physically relevant parameters are $\mu$ and $\rho$,
whereas, in the $|\Lambda_5|>0$ case, by defining $\Lambda_5\equiv
- 6/\ell^2$, we can see that, other than the $\mu$, the ratio
$\rho/\ell$ is more relevant (for determining $\Lambda_4$) than
$\rho$ or $\ell$ separately.

Third, from Eq.~(\ref{effective-S}),
 we find that the relation between four- and five-dimensional
 effective Planck masses is given by
\begin{equation}
 M_{Pl}^2 = {M_5^3\, \rho }\int_{-\infty}^{\infty} d{z}\,
 \frac{1}{\left(e^{\mu|z|}-c^2\, e^{-\mu|z|}\right)^3}\approx
 \frac{M_5^3 \rho }{4 \mu c^2}\left(\frac{1+c^2}{(1-c^2)^2}-
 \frac{\ln (1-c^2)}{2 c}\right),
\end{equation}
where $c^2\equiv (-\Lambda_5) (\rho^2/24\mu^2)$. In the limit $\mu
\to 0$, a new dimension of spacetime opens up, since $M_4^2 \to
\infty$, but with a nonzero $\mu$ (and a finite $c$) the 4D
effective Newton's constant is finite. All the cosmological
features described above extend far beyond 5D theory.

One basic reason for extending the discussion above to ten
dimensions~\cite{Ish:2010}, or more specifically, string theory,
is that it includes generalizations of both standard quantum field
theory and 4D Einstein's theory. If ``spacetime'' is an emergent
phenomenon, as suspected by many string theorists, then some of
the extra dimensions predicted by string theory can be realized
only when we probe sufficiently small distances. From this
viewpoint, it would be reasonable to assume that five of the extra
dimensions (predicted by string theory) are much smaller in size
than the radial (fifth) dimension. A few explicit examples were
provided in the Refs.~\cite{Ish:2010,Ish:2010b}, where explicit
solutions describing an accelerating expansion of the universe
were found just by solving the 10D Einstein equations.

\medskip

I conclude with a few remarks.\\
One of the most exciting developments in cosmology is the
suggestion that the warping of extra dimensions plays a key role
in determining the magnitude of the four-dimensional cosmological
constant and hence the rate of expansion of our physical universe.

The key point in the present method of explaining an accelerating
expansion of the universe is the treatment of `brane' as a
four-dimensional physical spacetime and consideration of a
non-factorizable (warped) background geometry in higher
dimensions. Within this set-up, the existence of an effective
cosmological constant-like term in four dimensions provides a
tantalizing possibility of explaining an accelerating expansion of
the universe from fundamental theories of gravity, which include
string theory.

String theory as it is currently formulated and interpreted has a
huge number of equally plausible solutions, called string
vacua~\cite{KKLT}. These vacua might be sufficiently diverse to
explain various phenomena we might observe at lower energies. If
these properties are true, string theory as a theory of quantum
gravity would have little predictive power for low energy particle
physics experiments and also for cosmologies. Because the theory
is difficult to test, some theoretical physicists have asked if it
can even be called a scientific theory of everything. In the book
{\it The Trouble with Physics}, Lee Smolin takes a complex debate
on topics such as string landscape and ability of string
compactificatons to explain the cosmological constant problem.
But, it is difficult to imagine that a completely wrong theory
could generate so many good ideas, including microscopic
descriptions of inflation from D-braneworld models and dualities
between closed string theories that contain gravity and decoupled
open string theories that don't.

Most importantly, we are not arguing that string theorists'
inability to empirically test their results in a cosmic lab will
disappear soon; shifting the emphasis from conventional
explanation to cosmic acceleration and/or cosmological constant
problem to a boarder picture of D-dimensional Einstein's theory,
with one or more warped extra dimensions, leads to a general class
of accelerating cosmologies which include in four dimensions
Einstein's gravity supplemented by an effective cosmological
constant-like term.

In the coming years we all hope to learn much more about inflation
and observed cosmic acceleration of the universe (attributed to a
small positive cosmological constant or other fluid-like ``dark
energy") from the highly refined computations and sophisticated
observations. More data from precision cosmology and large hadron
collider at CERN might help to address questions about the early
universe and the high energy frontiers. The years ahead will
certainly bring even more twists, breakthroughs and surprises in
gravity and cosmology research.

\section*{Acknowledgement}

This work was supported by the Marsden fund of the Royal Society
of New Zealand.



\begin{thebibliography}{99}
\itemsep 0pt

\bibitem{supernovae}
A. G. Riess {\it et al.}
  Astron.\ J.\  {\bf 116}, 1009 (1998);
  S. Perlmutter {\it et al.}, Astrophys. J. {\bf 517}, 565 (1999).


\bibitem{Online}
WMAP Project Webpage [http://map.gsfc.nasa.gov].

\bibitem{Polchinski}
  J.~Polchinski,
  Phys.\ Rev.\ Lett.\  {\bf 75}, 4724 (1995).

\bibitem{RS}
  L.~Randall and R.~Sundrum,
  Phys.\ Rev.\ Lett.\  {\bf 83}, 3370 (1999);
  L.~Randall and R.~Sundrum, 
  Phys.\ Rev.\ Lett.\  {\bf 83}, 4690 (1999).

\bibitem{Malda}
J. Maldacena,
Adv. Theor. Math. Phys. 2, 231 (1998).


\bibitem{KKLT}
S.~Kachru, R.~Kallosh, A.~Linde and S.~P.~Trivedi,
  Phys.\ Rev.\  D {\bf 68}, 046005 (2003).

\bibitem{Gibbons-84}
G.~W.~Gibbons, in {\it Supersymmetry, Supergravity and Related
Topics}, edited by F. del Aguila, J. A. de Azcarraga, and L. E.
Ibanz (World Scientific, 1985), pp. 123-146.

\bibitem{Malda-Nunez}
  J.~M.~Maldacena and C.~Nunez,
  Int.\ J.\ Mod.\ Phys.\  A {\bf 16}, 822 (2001).

\bibitem{GKP}
  S.~B.~Giddings, S.~Kachru and J.~Polchinski,
  Phys.\ Rev.\  D {\bf 66}, 106006 (2002).

\bibitem{TW}
  P.~K.~Townsend and M.~N.~R.~Wohlfarth,
  Phys.\ Rev.\ Lett.\  {\bf 91}, 061302 (2003);
N.~Ohta,
  Phys.\ Rev.\ Lett.\  {\bf 91}, 061303 (2003).

\bibitem{Ish03c}
I.~P.~Neupane,
  Class.\ Quant.\ Grav.\  {\bf 21}, 4383 (2004);
  I.~P.~Neupane and D. L.~Wiltshire,
  Phys.\ Rev.\  D {\bf 72}, 083509 (2005).

\bibitem{Ish07a}
I.~P.~Neupane,
  Phys.\ Rev.\ Lett.\  {\bf 98}, 061301 (2007).

\bibitem{Ish:2010}
  I.~P.~Neupane, Class. Quant. Grav. {\bf 26}, 195008 (2009);
Class.\ Quant.\ Grav.\  {\bf 27}, 045011 (2010);

\bibitem{Ish:2010b}
I.~P.~Neupane, Phys. Lett. B683, 88 (2010).

\end{thebibliography}
\end{document}